\documentstyle[12pt]{article}
\def\beq{\begin{equation}}
\def\eeq{\end{equation}}
\newcommand{\bea}{\begin{eqnarray}}
\newcommand{\eea}{\end{eqnarray}}
\newcommand{\nn}{\nonumber}
\def\noi{\noindent}
\font\boldmath=cmbsy10
\font\fourteenbf = cmbx10 at 14 pt
\mathchardef\mynabla="0235

\font\boldgreek=cmmib10
\mathchardef\mygamma="090D
\def\bfgamma{{\fam=9 \mygamma}\fam=1}
\mathchardef\mysigma="091B
\def\bfsigma{{\fam=9 \mysigma}\fam=1}
\mathchardef\mymu="0916 
\def\R{ {\rm R \kern -.31cm I \kern .15cm}}
\def\C{ {\rm C \kern -.15cm \vrule width.5pt \kern .12cm}}
\def\Z{ {\rm Z \kern -.27cm \angle \kern .02cm}}
\def\N{ {\rm N \kern -.26cm \vrule width.4pt \kern .10cm}}
\def\1{{\rm 1\mskip-4.5mu l} }

\begin{document}
\vbox to 1 truecm {}
\centerline{\bf NEW HEAVY QUARK LIMIT SUM RULES INVOLVING} \smallskip
\centerline{\bf ISGUR-WISE FUNCTIONS AND DECAY CONSTANTS}
\bigskip
\centerline{\bf A. Le Yaouanc, L. Oliver, O. P\`ene and J.-C. Raynal}
\centerline{Laboratoire de Physique Th\'eorique et Hautes
Energies\footnote{Laboratoire
associ\'e au Centre National de la Recherche Scientifique - URA D0063\\
\hspace*{\parindent} email: oliver@qcd.th.u-psud.fr, pene@qcd.th.u-psud.fr}}
\centerline{Universit\'e de Paris XI, b\^atiment 211, 91405 Orsay Cedex,
France}

\vskip 2 truecm
\noindent  {\bf \underbar{Abstract}.-} We consider the dominant $c\bar{c}$
contribution
to $\Delta \Gamma$ for the $B_s^0$-$\bar{B}_s^0$ system in the heavy quark
limit for
both $b$ and $c$ quarks. In analogy with the Bjorken-Isgur-Wise sum rule in
semileptonic
heavy hadron decay, we impose duality between the parton model calculation of
$\Delta
\Gamma$ and its estimation by a sum over heavy mesons. Varying the mass ratio
$m_c/m_b$
 and assuming factorization and saturation by narrow resonances $(N_c \to
\infty$), we
obtain new sum rules that involve the Isgur-Wise functions $\xi^{(n)}(w)$ and
$\tau^{(n)}_{1/2}(w)$ and the decay constants $f^{(n)}$, $f^{(n)}_{1/2}$ ($n$
stands for
any radial excitation). Alternatively, we deduce the sum rules with another
method free
of the factorization hypothesis, from the saturation of the expectation value
of a product
of two currents by heavy hadrons and by the corresponding free quarks. The sum
rules read
$\sum\limits_{n}{f^{(n)} \over f^{(0)}} \xi^{(n)}(w) = 2
\sum\limits_{n}{f^{(n)}_{1/2}
\over f^{(0)}} \tau^{(n)}_{1/2}(w) = 1$, valid for all $w$. Moreover, we
obtain, in the
heavy quark limit, $f^{(n)}_{3/2} = 0$. As a consequence, unlike the BIW sum
rule, the
slope of the elastic function $\xi (w)$ is related to radial excitations alone.
These are
generalizations, rigorous for QCD in the heavy quark limit, of results that
have an easy
understanding in the non-relativistic quark model.  \par

\vskip 1 truecm
\noindent LPTHE Orsay 96-28 \par
\noindent April 1996
\newpage
\pagestyle{plain}
The Bjorken-Isgur-Wise (BIW) \cite{[bjorken],[isgur]} sum rule can be obtained
from
duality between the estimation of a heavy hadron semileptonic decay rate by the
parton
model or by a sum over exclusive heavy hadrons, or equivalently from the
saturation of the
expectation value of a product of currents by heavy hadrons or by the
corresponding free
quarks. We can call this the duality approach, used in the paper by Isgur and
Wise
\cite{[isgur]}, where they estimate the hadronic tensor involved in the total
semileptonic
decay rate $W_{\mu \nu} = \sum\limits_{n}< B|J^{bc}_{\mu}|n > <
n|J^{cb}_{\nu}|B>$ either by the parton model, with the $c$ quark
con\-si\-dered as heavy,
or by a sum over heavy hadron intermediate states. Indeed, one can expect
duality owing
to the heavy quark limit, the corrections being subleading in $1/m_Q$, as it
has been
demonstrated in a well-defined formalism \cite{[shifman]}. This approach is to
be
distinguished from the current algebra approach, used by Bjorken et al.
\cite{[bjorken]}
that leads to the same result essentially because the two terms contributing to
the
commutator, the direct and $Z$-graph contributions, satisfy separately a sum
rule of
their own. \par
	To fix our ideas, let us write down the heavy meson current matrix elements in
 terms of
the corresponding Isgur-Wise (IW) functions. We have, for $S$-waves in terms of
the
function $\xi (w)$~:
\bea
&&< D({\rm v}_f)|V_{\mu}|\bar{B}({\rm v}_i) > = N \ \xi (w) \left ( {\rm v}_i +
{\rm v}_f
\right )_{\mu} \nn \\  &&< D^*({\rm v}_f,\varepsilon ) |V_{\mu}|\bar{B}({\rm
v}_i) > = i \
N \ \xi (w) \ \varepsilon_{\mu \alpha \beta \gamma} \  \varepsilon^{* \alpha} \
{\rm
v}_f^{\beta} \  {\rm v}_i^{\gamma} \nn \\ &&< D^*({\rm v}_f,\varepsilon
)|A_{\mu}|\bar{B}({\rm v}_i) > = N \ \xi (w) \left [ \left ( {\rm v}_i\cdot
{\rm v}_f
+ 1 \right ) \varepsilon^*_{\mu} - \left ( \varepsilon^* \cdot {\rm v}_i \right
) {\rm
v}_{f\mu} \right ]	\label{1}  \eea

\noindent where $N = {1 \over \sqrt{4{\rm v}^0_B {\rm v}_D^0}}$.
For the transition from $S$ to $P$-waves, there are two independent functions
$\tau_{1/2}(w)$, $\tau_{3/2}(w)$, where $j = {1 \over 2}$, ${3 \over 2}$  stand
for the
total angular momentum of the light degrees of freedom relative to the heavy
quark. We
list only the matrix elements that concern the specific purpose of this paper,
the
estimation of $\Delta \Gamma$, as will become clear below~:
\bea
&&< D( ^{1/2}1^+ ) ({\rm v}_f, \varepsilon )|V_{\mu}|\bar{B}({\rm v}_i) > = N \
2
\left [ (w - 1) \varepsilon_{\mu}^* - \left ( \varepsilon^* \cdot {\rm v}_i
\right )
{\rm v}_{f\mu} \right ] \tau_{1/2}(w) \nn \\
&&< D ( ^{1/2}1^+ ) ({\rm v}_f,\varepsilon ) |A_{\mu}|\bar{B}({\rm v}_i) > = N
\ i
\ \varepsilon_{\mu \alpha \beta \gamma} \
\varepsilon^{*\alpha}\left ( {\rm v}_i + {\rm v}_f \right )^{\beta} \left (
{\rm v}_i -
{\rm v}_f \right)^{\gamma} \tau_{1/2}(w) \nn \\
&&< D ( ^{1/2}0^+ ) ({\rm v}_f)|A_{\mu}|\bar{B}({\rm v}_i) > = N \ 2
\left ( {\rm v}_f - {\rm v}_i \right )_{\mu} \tau_{1/2}(w)	\ \ \ . \label{2}
\eea

Identifying the parton model $B$ meson total semileptonic decay rate to its
estimation by
the sum over the possible meson final states listed above, one finds the BIW
sum rule~:
\beq
\sum_{n} {w + 1 \over 2} \left \{ \left [ \xi^{(n)}(w) \right ]^2 +
(w - 1) \left [ 2 \left [ \tau^{(n)}_{1/2}(w) \right ]^2 +
(w + 1)^2 \left [ \tau^{(n)}_{3/2}(w) \right ]^2 \right ] + ... \right \} = 1 \
. \label{3}
\eeq
\noindent Where $n$ stands for radial excitations and the dots indicate a
possible continuum. For the transition to a radial excitation one has
$\xi^{(n)}(w) \sim
(w - 1)$ ($n \not= 1$) due to the orthogonality of the wave functions at zero
recoil.
Notice that for these transitions we have changed the notation relatively to
Isgur and
Wise \cite{[isgur]}, as in their paper they define $\xi^{(n)}(w)$ factorizing
out
$(w - 1)$. It is more convenient for us to keep an
homogeneous definition for any $S$-wave meson in the final state. It is well
known that
from (\ref{3}) follows the sum rule for the slope of the ground state IW
function, and
Bjorken's lower bound for it~:
\beq
\rho^2 = {1 \over 4}  + \sum_{n} \left \{ \left [
\tau^{(n)}_{1/2}(1) \right ]^2 + 2 \left [ \tau^{(n)}_{3/2}(1) \right ]^2
\right \}
\ \ \ , \qquad		\rho^2 \geq  {1 \over 4}   \label{4}
\eeq

We want now to apply duality to a non-leptonic process, $\Delta \Gamma$, the
decay rate
difference between the two CP eigenstates of the system $B^0_s$-$\bar{B}^0_s$
(we
assume CP to be conserved). Our aim is to obtain, in the factorization
approximation, sum rules involving both annihilation constants and form
factors.\par

Before proceeding to the calculation, we need to make a number of important
remarks
on questions of principle. \par

First, there is an essential difference between the total semileptonic rate $B
\to X_c
\ \ell \nu$  and $\Delta \Gamma$. In the former, by varying $q^2$, the
(mass)$^2$ of the lepton system, one has equi\-valently a variable $w$ for the
hadronic matrix element $<X_c|J_{\mu}|B>$. On the contrary, $\Delta \Gamma$ is
a
non-leptonic quantity, and, for a given intermediate state, $q^2$ is
fixed. To set our ideas, let us remind that there are two contributions to
$\Delta
\Gamma$, corresponding to two ways of cutting the box diagram~: the exchange
contribution
(Fig.~1) and the spectator contribution (Fig.~2), as pointed out by Hagelin in
his
pioneering work within the parton model \cite{[hagelin]} and in
ref.~\cite{[aleksan]},
where we discussed $\Delta \Gamma$ from the point of view of ground state
exclusive
modes. As will be justified later, we will only consider the spectator
contribution.
 For the spectator contribution, one has two types of color singlet groupings,
$(c\bar{s})_1(\bar{c}s)_1(D_s \bar{D}_s \ldots )$ and $(c
\bar{c})_1(s\bar{s})_1
(\psi \varphi \ldots)$. As we will argue below, we will only consider the
former. We draw
a typical contribution of this type in Fig.~3. It is clear that any $B \to D$
form factor
$F^{DB}(q^2)$ ($D$ is any meson with charm quantum number) must be taken at
$q^2 \cong
m^2_c$  because there is emission of another $D$ meson, and the charmed quark
is heavy.
In the heavy quark limit, a form factor $F^{DB}(q^2) \to \eta (w)$ (up to
kinematic
factors), where $\eta (w)$ is a generic IW function. At $q^2 = m^2_c$ the
variable $w =
{m^2_b + m^2_c - q^2 \over 2m_b m_c}$  will take a fixed value \beq q^2 = m^2_c
	\to 	w =
{m_b \over 2m_c}		 	\label{5} \eeq However, without any assumption on the value
of $m_c$,
$m_b$ (except to be heavy), by varying the mass ratio ${m_b \over m_c}$ one can
express $\Delta \Gamma$ in terms of $w$, both in the parton model and in the
exclusive calculation as well. The parton model rate difference will be given
in terms of
some kinematic function of $w$, while the exclusive calculation will be
expressed in terms of the IW functions $\xi^{(n)}(w)$, $\tau^{(n)}_{1/2}(w)$
and $\tau^{(n)}_{3/2}(w)$. \par

The exclusive calculation of $\Delta \Gamma$ involves also other quantities,
namely, within the factorization approximation, the
following decay constants~:
\bea
&&< D_s(0^-)({\rm v})|A^{sc}_{\mu}|0> = N \ f_D \sqrt{m_D} \ {\rm v}_{\mu} \nn
\\
&&< D_s(1^-)({\rm v}, \varepsilon ) |V^{sc}_{\mu}|0 > = N \ f_D \sqrt{m_D}
\ \varepsilon_{\mu}^*			\label{6}
\eea
\noindent for $S$ states, where $N = {1 \over \sqrt{2{\rm v}_D^0}}$, and
\bea
&&< D_s ( ^{1/2}0^+  )({\rm v})|V^{sc}_{\mu}|0> = N
\ f^{(1/2)}_D \ \sqrt{m_D} \ {\rm v}_{\mu} \nn \\
&&< D_s  ( ^{1/2}1^+  ) ({\rm v},\varepsilon ) |A^{sc}_{\mu}|0 > = N
\ g^{(1/2)}_D \sqrt{m_D} \  \varepsilon^*_{\mu} \nn \\
&&< D_s ( ^{3/2}0^+ ) ({\rm v},\varepsilon )|A^{sc}_{\mu}|0 > = N
\ g^{(3/2)}_D \sqrt{m_D} \  \varepsilon^*_{\mu} \label{7}
\eea
\noindent for $P$ states. The annihilation constant $f^{(1/2)}$ is non-zero
because the
vector current is not conserved. The annihilation constants are defined in such
a way
that they are flavor-independent in the heavy quark limit. \par

We will identify ${m_b \over 2m_c} \equiv w$ in the parton model and in the
exclusive
calculation. By imposing duality, we will then obtain sum rules involving IW
functions and decay cons\-tants. We will begin by considering the physical V-A
case (the
ground state contribution to $\Delta \Gamma$ and duality as $2m_c \to m_b$ were
studied
in \cite{[aleksan]}), but we must keep in mind that the heavy quark limit
concerns QCD alone
and is independent of the Lorentz structure of the current. As we will see
below, we
will obtain various sum rules by varying the chirality structure of the
electroweak part
of the theory. Indeed, we shall use $\Delta \Gamma$ as an intermediary step to
get a
sum rule for annihilation constants and form factors. For this purpose we can
choose a
fictitious but convenient four fermion interaction. Concerning short distance
QCD
coefficients we can choose in particular $c_+ = c_- = 1$.  \par

In the following, we will assume that the excitation energies verify the
hierarchy
$\Delta E_n \ll p = \left ( {1 \over 4} m_b^2 - m_c^2 \right )^{1/2}$, and
$\Delta E_n
\ll m_c$, $m_b$. Strictly speaking, the excitation energies range up to
infinity and
the latter hierarchy cannot be valid for all states. In practice we assume that
these
higher states do not contribute significantly in the sum rules. One could
introduce a
cut-off $\mu$ as in \cite{[isgur]}, but we preferred to avoid any unessential
complication. On the other hand, we will neglect hard gluons, because
these appear as corrections to the heavy quark limit. \par

To be able to get sum rules involving IW functions and decay constants
(semileptonic
quantities) from  a non-leptonic quantity like $\Delta \Gamma$ it would seem
that we need
to be in the limit in which factorization (i.e. vacuum insertion) holds for the
non-leptonic decays, i.e. the $N_c \to \infty$ limit. \par

Taking $c_+ = c_- = 1$, the amplitude to
decay into the color grouping $(c\bar{c})_1(s\bar{s})_1$ is proportional to
$1/N_c$. As
pointed out by Shifman \cite{[shifman2]}, this type of final state violates
duality at
$O(1/N_c)$. However, we have shown that the sum of both color configurations
$(c\bar{s})_1(\bar{c}s)_1$ and $(c\bar{c})_1 (s\bar{s})_1$ satisfy duality if
the
final state interaction (FSI) is taken into account \cite{[leyaouanc]}. But
since we want here to relate semileptonic quantities and we therefore adopt
$N_c \to
\infty$ to avoid violations to factorization, we take, to summarize~: $c_+ =
c_- = 1$,
$N_c \to \infty$. Therefore, factorization holds, the color configurations
$(c\bar{c})_1(s\bar{s})_1$ vanish and, moreover, the width is saturated by
narrow
resonances. \par

The estimation of the exchange contribution to $\Delta \Gamma$ in the
parton model is straightforward \cite{[hagelin]}. However, the calculation
using exclusive
channels would involve form factors $F^{DD}(m^2_b)$, quantities that are very
different
\cite{[aleksan]} from the ones we are interested in the case of the
spectator contributions, namely $F^{DB}(m^2_c)$. Therefore, we need to compare
the
contributions $\Delta \Gamma^{partons}_{spectator}$ to $\Delta
\Gamma^{exclusive}_{spectator}$ alone. \par

However, let us proceed with care and make explicit the $N_c$ dependence. Let
us substract
from the parton model result
\beq
\Delta \Gamma^{partons} = {G^2 \over 2\pi} f_B^2 |V^*_{cb} V_{cs}|^2 p
\label{8}
\eeq
\[ \left \{ 2 \left ( 1 -
{1 \over 2N_c} \right ) 2m_b^2 \left [ {1 \over 2} - {1 \over 3} \left ( {3
\over 4}  +
{p^2 \over m_b^2} \right ) \right ] - 2 \left ( 1 + {1
\over N_c} \right ) \left [ m_c^2 - {1 \over 2} m_b^2 +
{1 \over 3} m_b^2 \left ( {3 \over 4}  +
{p^2 \over m_b^2} \right ) \right ] \right \} \]
\noindent (the two terms correspond respectively to the two operators
contributing to the
mixing $\left [ \bar{c} \gamma_{\mu} \left ( 1 - \gamma_5 \right )
b \right ] \left [\bar{c} \gamma_{\mu} \left ( 1 - \gamma_5 \right ) b \right
]$ and
$\left [ \bar{c} \left ( 1 + \gamma_5 \right ) b \right ] \left [ \bar{c} \left
(
1 + \gamma_5 \right )b \right ]$) the exchange contribution~:
\beq
\Delta \Gamma^{partons}_{exchange} = - {G^2 \over 2\pi}\ {2 \over N_c}
\ f_B^2 \ m_c^2 |V^*_{cb} V_{cs}|^2 p		\label{9}
\eeq
\noindent where $p = \left ( {1 \over 4} m_b^2 -
m_c^2 \right )^{1/2}$ is the three-momentum. We get the simple result
\beq
\Delta \Gamma^{partons}_{spectator} = {G^2 \over 2\pi} \ f_B^2 \ m_b^2
|V^*_{cb}
V_{cs}|^2 p		\ \ \ . 		\label{10}
\eeq

 We need now to compute the sum
over exclusive modes. Let us see what are the constraints on the annihilation
constants
defined in (\ref{7}). Let us first realize that, on grounds of Lorentz
covariance and
parity,
\bea < D_s ( ^{1/2}0^+ ) ({\rm v})|A^{sc}_{\mu}|0> &=& <D_s ( ^{3/2}2^+ )
({\rm v},\varepsilon )|A^{sc}_{\mu}|0 > \nn \\  &=& < D_s ( ^{3/2} 2^+ ) ({\rm
v},
\varepsilon )|V^{sc}_{\mu}|0 > \ \equiv 0	\ \ \ . \label{11}  \eea
 \noindent On the other hand, using (for the longitudinal polarizations of
$1^+$,
$2^+$),  \bea
&&S^3_c|D_s  ( ^{1/2}1^+  )({\bf 0},0) > = - {1 \over 2}|D_s  (
^{1/2}0^+  ) ({\bf 0}) > \nn \\
&&S^3_c|D_s  ( ^{3/2}1^+  )({\bf 0},0) > = - {1 \over
2}|D_s  ( ^{3/2}2^+  ) ({\bf 0},{\bf 0}) >		 \label{12}
\eea
\noindent and the commutation relations \cite{[aleksan]}
\beq
\left [ S^3_c , A^{sc}_3 \right ] = - {1 \over 2} V^{sc}_0 \qquad \qquad
\left	[
S^3_{c} , V^{sc}_0 \right ] = - {1 \over 2} A^{sc}_3 			\label{13}
\eeq
\noindent together with (\ref{11}), we obtain~:
\beq
f^{(1/2)} = g^{(1/2)} 	\qquad g^{(3/2)} = 0 	\qquad \hbox{(heavy mass limit)} \
\ \ .
\label{14}
\eeq
This result of the heavy quark limit (one heavy quark and one light quark) is
quite
different from the usual intuition of the case of mesons made of
quark-antiquark of equal
masses, in which the state $1^{+-}$ is decoupled of the axial current, and the
state
$0^{++}$ is decoupled from the vector current, owing to current conservation.
These
decouplings are completely general in QCD for mesons with equal mass valence
quarks. In
the language of the quark model, in the equal mass limit, from the change of
basis \bea
&&|^1P_1 > = \scriptstyle{\sqrt{{1 \over 3}}} \ |^{1/2}1^+ > +
\scriptstyle{\sqrt{{2 \over
3}}} \ |^{3/2}1^+ > \nn \\ &&|^3P_1 > = - \scriptstyle{\sqrt{{2 \over 3}}} \
|^{1/2}1^+ >
+ \scriptstyle{\sqrt{{1 \over 3}}} \ |^{3/2}1^+ >			 \label{15}  \eea
\noindent (we neglect the {\bf L.S} coupling) one gets indeed
\beq
g^{(1/2)}  + \sqrt{2} \ g^{(3/2)}  = 0 	\qquad f^{(1/2)} = 0 	\qquad
\hbox{(equal mass
limit)}	 \ \ \ . \label{16}
\eeq
\noindent One can understand intuitively the transition between these two
regimes by using
the non-relativistic quark model with unequal masses. In the lowest order in
the quark
velocities, one has, in terms of the relative momentum within the bound state~:
\beq
V^{sc}_0  \to u_c^+ v_s  \to {1 \over 2\sqrt{2}} \left ( {1 \over m_c} - {1
\over
m_s} \right) \chi_c^+ \ \bfsigma \cdot {\bf p} \ \chi_s 				\label{17}
\eeq
\[ A^{sc}_3  \to u_c^+ \sigma_3 v_s \to {1 \over 2\sqrt{2}} \left ( {1 \over
m_c} -
{1 \over m_s} \right ) \chi_c^+ \ p_z \ \chi_s + {1 \over 2 \sqrt{2}} \left (
{1 \over
m_c} + {1 \over m_s} \right ) \chi_c^+ \ i( \bfsigma \times {\bf p})_z  \
\chi_s \ . \]
\noindent After some angular momentum calculations, one obtains (\ref{14}) in
the $m_c \to
\infty$ limit, and (\ref{16}) in the equal mass limit. \par

An important consequence of (\ref{14}) is that, in the problem that we are
considering of
the estimation of $\Delta \Gamma$, the emission of $^{3/2}1^+$ states is
forbidden in the
hypothesis of factorization. \par

$\Delta \Gamma$ will read, in the estimation through exclusive modes~:
\[ \Delta \Gamma^{exclusive}_{spectator} = 2\Gamma_{12} = {G^2 \over 4\pi} \
m_B |V^*_{cb} V_{cs}|^2 p \]
\[ \left \{ \sum_{n,n'} <0|V^{sc}_{\mu} - A^{sc}_{\mu}|D_n({\rm v}_f)>
<B^0({\rm v}_i)|V^{cb}_{\mu} - A^{cb}_{\mu}|\bar{D}_{n'}({\rm v}'_f)> \right .
\]
\beq
\left .	\times <
\bar{D}_{n'}({\rm v}'_f)|V^{sc}_{\alpha} - A^{sc}_{\alpha}|0 > <
D_n({\rm v}_f)|V^{cb}_{\alpha} - A^{cb}_{\alpha}|\bar{B}^0({\rm v}_i) > \right
\}
\label{18} \eeq
\noindent where the factor $p$ is the three-momentum that also appears in the
parton model
expression (\ref{10}). For the sake of clarity, it is convenient to split the
different
contributions into three types (we call from now on the exclusive spectator
contribution
simply $\Delta \Gamma$)~:
\beq
\Delta \Gamma = \Delta \Gamma^{(- -)} + \Delta \Gamma^{(- +)} + \Delta
\Gamma^{(+
+)}	\label{19}
\eeq
\noindent where $\Delta \Gamma^{(- -)}$, $\Delta \Gamma^{(- +)}$ and $\Delta
\Gamma^{(+
+)}$ denote respectively two $S$-wave mesons of parity - (the ground state and
its radial
excitations), one $S$-wave meson with parity - and one $P$-wave meson with
parity +, and
two $P$-wave mesons of parity + in the final state \cite{[note]}. \par

Since each transition matrix element is contracted with a creation matrix
element out of
the vacuum, we need a number of current matrix elements contracted with
four-vectors.
Recall the identification ${m_b \over 2m_c} \equiv w$. \par

1) Transition matrix element to $S$-wave meson contracted with pseudoscalar or
scalar
emission (since the velocity of the emitted meson is $(m_B {\rm v}_i - m_D
{\rm v}_f)/m_D)$~:
\newpage
 \[ {1 \over m_D} < D({\rm v}_f)|\left  ( m_B{\rm v}_i - m_D {\rm v}_f
\right ) \cdot V|\bar{B}({\rm v}_i) > = N \ \xi (w) \ (2w - 1) (w + 1) \]
\[{1 \over m_D} < D^*({\rm v}_f, \varepsilon ) |(m_B{\rm v}_i - m_D{\rm v}_f)
\cdot
V|\bar{B}({\rm v}_i) > =\] \[ = i \ N \ \xi (w) {1 \over m_D} \varepsilon_{\mu
\alpha
\beta \gamma} \left ( m_B {\rm v}_i - m_D {\rm v}_f \right )^{\mu} \
\varepsilon^{*\alpha}
\ {\rm v}_f^{\beta} \ {\rm v}_i^{\gamma}) \equiv 0 \]
\beq
{1 \over m_D} < D^*({\rm v}_f, \varepsilon )|\left ( m_B {\rm v}_i - m_D{\rm
v}_f \right )
\cdot A|\bar{B}({\rm v}_i) > = N \ \xi (w) \sqrt{w^2 - 1} (2w + 1)	 \label{20}
\eeq

2) Transition matrix element to $S$-wave meson contracted with vector or axial
emission
($\lambda$ indicates the polarization)~:
\bea
&&< D({\rm v}_f)|\varepsilon '^* \cdot V|\bar{B}({\rm v}_i) > = - N \ \xi (w)
(2w + 1)
\sqrt{w^2 - 1} \nn \\
&&< D^*({\rm v}_f, \varepsilon )|\varepsilon '^* \cdot V|\bar{B}({\rm v}_i) > =
\pm i \
N \ \xi (w) \sqrt{w^2 - 1} \qquad (\lambda = \pm 1, \lambda ' = \mp 1) \nn \\
&&< D^*({\rm v}_f, \varepsilon )|\varepsilon '^* \cdot A|\bar{B}({\rm v}_i) > =
- N \ \xi
(w) (w + 1) (2w - 1)	\qquad (\lambda = 0) \nn \\
&&< D^*({\rm v}_f,\varepsilon )|\varepsilon '^* \cdot A|\bar{B}({\rm v}_i) > =
- N \ \xi
(w) (w + 1)		 \qquad (\lambda = \pm1 , \lambda ' = \pm 1) \label{21}
\eea

3) Transition matrix element to $P$-wave meson contracted with pseudoscalar or
scalar
emission~:
\bea
&&{1 \over m_D} < D \left ( ^{1/2}1^+ \right ) ({\rm v}_f,\varepsilon
) | \left ( m_B {\rm v}_i - m_D {\rm v}_f \right ) \cdot V |\bar{B}({\rm v}_i)
> = - N \ 2
\ \sqrt{w^2 - 1} (2w - 1) \tau_{1/2}(w) \nn \\
&&{1 \over m_D} < D \left ( ^{1/2}1^+ \right ) ({\rm v}_f,\varepsilon )|\left (
m_B{\rm v}_i - m_D {\rm v}_f \right )\cdot A|\bar{B}({\rm v}_i) > \equiv 0
\label{22}
\nn \\ &&{1 \over m_D} < D \left ( ^{1/2} 0^+ \right ) ({\rm v}_f)|\left ( m_B
{\rm v}_i -
m_D {\rm v}_f \right ) \cdot A|\bar{B}({\rm v}_i) > = N \ 2 \ (2w + 1) (w - 1)
\
\tau_{1/2}(w)   \eea

4) Transition matrix element to $P$-wave meson contracted with vector or axial
emission~:
\bea
&&< D  ( ^{1/2}1^+  ) ({\rm v}_f , \varepsilon )| \varepsilon '^{*} \cdot
V|\bar{B}({\rm v}_i) > = N \ 2(w - 1) (2w + 1) \ \tau_{1/2}(w) \quad (\lambda =
0) \nn \\
&&< D  ( ^{1/2}1^+  ) ({\rm v}_f, \varepsilon )|\varepsilon '^{*} \cdot
V|\bar{B}({\rm v}_i) > = - N \ 2(w - 1) \ \tau_{1/2}(w)	\quad (\lambda =  \pm
1, \lambda
' = \pm 1) \nn \\ &&< D  ( ^{1/2}1^+ ) ({\rm v}_f, \varepsilon ) |\varepsilon
'^{*} \cdot
A|\bar{B}({\rm v}_i) > = \pm i \ N \ 2 \ \sqrt{w^2 - 1} \ \tau_{1/2}(w) \quad
(\lambda =
\pm 1, \lambda ' = \mp 1) \nn \\ 		 &&< D  ( ^{1/2} 0^+ ) ({\rm v}_f) |
\varepsilon '^{*}
\cdot A | \bar{B} ({\rm v}_i)> = - N \ 2 (2w - 1) \sqrt{w^2 - 1} \
\tau_{1/2}(w) \
(\lambda = 0 ) \ . \label{23}
\eea
\noindent When summing the different contributions one must take special care
in their
relative sign. Let us detail, as an example, the calculation of the sum over
the ground
state mesons. From (\ref{20}) and (\ref{21}) we get, for the ground state
contributions~:
\newpage
\beq
\Delta \Gamma^{ground \ state} = {G^2 \over 8\pi}  f^2_B
m^2_B |V^*_{cb} V_{cs}|^2 p \left [ \xi (w) \right ]^2 {1 \over 4w^2}
\label{24}
\eeq
\[ \left \{ (2w - 1)^2 (w + 1)^2 + 2(w^2 - 1) (2w + 1)^2 + (w + 1)^2 (2w - 1)^2
+
2(w + 1)^2 - 2(w^2 - 1) \right \} \]
\noindent where the terms correspond respectively to $PP$, $PV$, $VV(p.v.L)$,
$VV(p.v.T)$, $VV(p.c.T)$ ($P$, $V$ stand for pseudoscalar, vector~; $p.v.$,
$p.c.$ for
parity violating and parity conserving~; $L$, $T$ for longitudinal,
transverse). We
recover the same signs for the different states that we found in
ref.~\cite{[aleksan]}, in
particular for the non-trivial case of the $PV$ contribution. We get therefore
:
\beq
\Delta \Gamma^{ground \ state} = {G^2 \over 2\pi}  f^2_B m^2_B |V^*_{cb}
V_{cs}|^2 p
\ {(2w - 1)^2 (w + 1)^2 \over 4w^2} [ \xi(w)]^2	\ \ \ .\label{25}
\eeq
\noindent Notice that in (\ref{25}) we have adopted the heavy quark limit
relation~:
\beq
f^2_B m_B  =  f^2_D m_D \label{26}
\eeq
\noindent in order to have the same overall factors as in the parton model
expression
(\ref{10}). \par

The sum over all excitations is simple in its final expression. To this aim,
let us define
the scale invariant sums extended over all radial excitations~:
\bea
&&X(w) = \sum_{n} {f^{(n)} \over f^{(0)}} \ \xi^{(n)}(w) \nn \\
&&T_{1/2}(w) = \sum_{n}{f^{(n)}_{1/2} \over f^{(0)}}
\ \tau^{(n)}_{1/2}(w)	\label{27}
\eea
where $f^{(0)}$ denotes the ground state annihilation constant. Obviously, one
has
\beq
X(1) = 1 \label{28bis}
\eeq

 From (\ref{25}) we can write down the sum over all $S$-wave radial
excitations, as defined
in (\ref{19})~:
\beq
\Delta \Gamma^{(- -)} = {G^2 \over 2\pi}  f^2_B m^2_B |V^*_{cb} V_{cs}|^2 p \
{(2w -
1)^2 (w + 1)^2 \over 4w^2} [X(w)]^2 \ \ \ . \label{28}
\eeq
\noindent The calculation of the $P$-wave meson contributions $\Delta
\Gamma^{(- +)}$ and
$\Delta \Gamma^{(+ +)}$ can be done using formulae (\ref{20})-(\ref{23}) and is
rather
tedious. One must be careful with the relative signs. Using the
notation (\ref{27}) we find
\bea
&&\Delta \Gamma^{(- +)} = - {G^2 \over 2\pi}  f^2_B m^2_B |V^*_{cb}V_{cs}|^2 p
\ {(4w^2 -
1) (w^2 - 1) \over 4w^2} 2[X(w)] \left [ 2T_{1/2}(w) \right ] \nn \\
&&\Delta \Gamma^{(+ +)} = {G^2 \over 2\pi}  f^2_B
m^2_B |V^*_{cb} V_{cs}|^2 p \ {(2w + 1)^2 (w - 1)^2 \over 4w^2}
\left [ 2T_{1/2}(w) \right ]^2 \ \ \ . 	\label{29}
\eea
\noindent Therefore, we obtain for the total sum~:
\bea
&&\Delta \Gamma = {G^2 \over 2\pi}  f^2_B m^2_B V^*_{cb} V_{cs}|^2 p \nn \\
&&{1 \over 4w^2} \left \{ (2w^2 - 1) \left [ X(w)-2T_{1/2}(w) \right ] + w
\left [ X(w) + 2T_{1/2}(w) \right ] \right \}^2	 \ \ \ . \label{30}
\eea
\noindent Identifying with the parton model calculation (\ref{10})~:
\beq
{1 \over 4w^2} \left \{ (2w - 1) (w + 1) X(w) - 2(2w + 1) (w - 1) T_{1/2}(w)
\right
\} ^2 = 1	 \ \ \ . \label{31}
\eeq
\noindent This equation is not the only one that one can obtain from duality
arguments for
$\Delta \Gamma$ in the heavy quark limit. As pointed out above, the heavy quark
symmetry
limit concerns QCD and is independent of the structure of the electroweak
theory. We can
obtain other relations of the type (\ref{31}) by just varying the chirality
structure of
the theory. \par

For example if instead of having the V-A structure $\gamma_{\mu}L$ one would
have a
pure vector coupling $\gamma_{\mu}$, one would select just a few states in the
exclusive calculation. Only production and annihilation by a vector current can
occur,
and then only the parity conserving vector-vector final state, i.e. $D^*
\bar{D}^*$ in the
relative $P$ wave (and its radial excitations) contribute. In particular, there
is no
contribution of the $P$ states~! The calculation follows along similar lines as
the case
of the V-A current. We find, in the $N_c \to \infty$ limit~: \beq
\Delta \Gamma^{partons}_{spectator}(pure \ V) = - {G^2 \over 4\pi} \ f_B^2 \
m_b^2
|V^*_{cb} V_{cs}|^2 \ p \ {w^2 - 1 \over 2w^2} 			\label{32}
\eeq
\beq
\Delta \Gamma^{exclusive}_{spectator}(pure \ V) = - {G^2 \over 4\pi} \ f_B^2 \
m_b^2
|V^*_{cb} V_{cs}|^2 \ p \ [X(w)]^2 \ {w^2 - 1 \over 2w^2} \label{33}
\eeq
\noindent Duality implies, in this case~:
\beq
[X(w)]^2 = 1			\ \ \ . \label{34}
\eeq
One could consider also a pure axial coupling $\gamma_{\mu}\gamma_5$. Then,
only
production and annihilation by an axial current can occur, and only the parity
conserving
axial-axial states $D^{**}(1^+)\bar{D}^{**}(1^+)$ in the relative $P$ wave (and
its radial
excitations) contribute. In particular, there is no contribution of the ground
state. We
find, in the $N_c \to \infty$ limit, the same result for the parton model
calculation~:
\beq
\Delta \Gamma^{partons}_{spectator}(pure \ A) = - {G^2 \over 4\pi} f_B^2 \
m_b^2
|V^*_{cb}V_{cs}|^2 \ p  \ {w^2 - 1 \over 2w^2} \label{35}
\eeq
\noindent and, for the exclusive calculation~:
\beq
\Delta \Gamma^{exclusive}_{spectator}(pure \ A) = - {G^2 \over 4\pi}
f_B^2 \ m_b^2 |V^*_{cb} V_{cs}|^2 \ p \ \left [ 2T_{1/2}(w) \right ]^2
\ {w^2 - 1 \over 2w^2} 	\ \ \ . \label{36}
\eeq
\noindent And duality implies~:
\beq
\left [ 2T_{1/2}(w) \right ]^2 = 1		\ \ \ . \label{37}
\eeq
{}From (\ref{31}), (\ref{34}) and (\ref{37}), owing to the fact that $X(w)$ and
$T_{1/2}(w)$
are real and we have the normalization condition (\ref{28bis}), we obtain the
simple
sum rules, the main result of this paper~:		 \beq
X(w) = 2T_{1/2}(w) = 1		\ \ \ . \label{38}
\eeq
\noindent This solution is very constrained, since $X(w)$ and $T_{1/2}(w)$ are
functions
of $w$. \par

Let us now use another method to obtain this result, namely duality applied to
a
product of currents $J^{sc}(x)$ $J^{cb}(y)$, as indicated in Fig. 4. This
method is
simpler as it gives a linear equation involving $X(w)$ and $T_{1/2}(w)$ that
leads
precisely to (\ref{38}). Moreover, the method does not rely on the hypothesis
of
factorization when we have used $\Delta\Gamma$ ($N_c \to \infty$). It seems to
have a
higher degree of generality. Let us consider the currents : \beq
J^{sc}(x) = \bar{s}(x) \ \Gamma^{sc} \ c(x)		\qquad J^{cb}(x) = \bar{c}(x) \
\Gamma^{cb} \ b(x)	\ \ \ . \label{39}
\eeq
\noindent We will estimate the matrix element
\beq
< 0|J^{sc}(0) \ \widetilde{J}^{cb}({\bf q})|\bar{B}_s({\rm v}_B) > 	\label{40}
\eeq
\noindent within the parton model approach in the heavy mass limit for both $c$
and
$b$ quarks and, on the other hand, by taking heavy hadrons as intermediate
states. We will first compute the product of currents $J^{sc}(0)$
$\widetilde{J}^{cb}({\bf q})$ acting on a $b$ quark state $|b, p_b, s_b >$
where
$\widetilde{J}^{cb}({\bf q})$ is the Fourier transform~: \beq
\widetilde{J}^{cb}({\bf q})
= \int d{\bf x} \ J^{cb}({\bf x},0) \ e^{-i{\bf q}\cdot {\bf x}} 	\ \ \ .
\label{41}
\eeq
We will have, in the heavy mass limit for the $c$ and $b$ quarks, as one
replaces the
heavy quark propagator by the free propagator~: \[J^{sc}(0) \
\widetilde{J}^{cb}({\bf q})
|b, p_b , s_b > =\] \beq
\bar{s}(0) \int d{\bf x} \ e^{-i {\bf q} \cdot {\bf x}} \left [ <0|c(0) \bar{c}
({\bf x}
, 0)|0> \right ]_{E > 0} \ b({\bf x}, 0)|b , p_b , s_b > \label{42}
\eeq
\noindent where
\[\left [ < 0|c(0) \bar{c}({\bf x},0)|0 > \right ]_{E > 0} = \int {d{\bf p}
\over
(2\pi)^3} \ {{/ \hskip - 2 mm}{{\rm v}}_p +1 \over 2{\rm v}^0_p} e^{-i{\bf
p}\cdot {\bf
x}}\]  \beq
b({\bf x},0)|b, p_b, s_b > = e^{-i{\bf p}_b \cdot {\bf x}}\  b(0)|b, p_b, s_b >
\ \
\ ,			 \label{43}
\eeq
\noindent with ${\rm v}_p^{\mu} = {p^{\mu} \over m_c}$. Finally,
\beq
J^{sc}(0) \ \widetilde{J}^{cb}({\bf q})|b, p_b, s_b > = \bar{s}(0) \left [
\Gamma_{sc}
{{/ \hskip - 2 mm}{{\rm v}}_c + 1 \over 2{\rm v}_c^0} \Gamma^{cb} \right ] b(0)
\label{44} \eeq
\noindent where $p_c = p_b - q$. Particularizing to the product of currents
$A^0V^0$, i.e.
$\Gamma^{sc} = \gamma^0 \gamma_5$ and $\Gamma^{cb} = \gamma^0$ we obtain~:
\beq
\gamma^0 \gamma_5 {{/ \hskip - 2 mm}{\rm v}_c +1 \over 2{\rm v}^0_c} \gamma^0 =
{{\rm v}^0_c \gamma^0 \gamma_5 - \gamma_5 + {\bf v}_c \cdot \bfgamma \gamma_5
\over 2{\rm v}^0_c}				 \ \ \ . \label{45}
\eeq
\noindent Using $< 0|\bar{s} \gamma^{\mu} \gamma_5 b|B({\rm v}_B) > = -
{\rm v}_B^{\mu} < 0|\bar{s} \gamma_5 b|B({\rm v}_B) > = {f_B \sqrt{m_B}
{\rm v}_B^{\mu} \over \sqrt{2{\rm v}^0_B}}$ we obtain, in the limit ${\rm v}_c
=
{\rm v}_D$, ${\rm v}_b = {\rm v}_B$~: \bea
< 0|J^{sc}(0) \ \widetilde{J}^{cb}({\bf q})|B({\rm v}_B) > &=& < 0|\bar{s}(0)
\gamma^0
\gamma_5 {{/ \hskip - 2 mm}{{\rm v}}_c + 1 \over 2{\rm v}^0_c} \gamma^0 \
b(0)|\bar{B}_s({\rm v}_B) > \nn \\  &=& {f_B \sqrt{m_B} \over 2{\rm v}^0_D
\sqrt{2{\rm v}^0_B}}  \left ( {\rm v}^0_B \ {\rm v}^0_D + 1 + {\bf v}_D \cdot
{\bf v}_D
\right ) \ \ \ . \label{46} \eea
\noindent We must now compute the matrix element (\ref{40}) by a sum over
exclusive heavy
hadrons $|n>$ as intermediate states. Moreover, we will assume dominance of the
sum
by intermediate narrow resonances. Using the formulae above for transition and
annihilation matrix elements, we find~:
\[< 0|J^{sc}(0) \ \widetilde{J}^{cb}({\bf q})|B({\rm v}_B) > = {f_B
\sqrt{m_B} \over 2{\rm v}^0_D \sqrt{2{\rm v}^0_B}}\]
\beq
\left [ X(w) \ {\rm v}^0_D \left ( {\rm v}^0_D + {\rm v}^0_B \right ) +
2T_{1/2}(w)
{\bf v}_D \cdot \left ( {\bf v}_B - {\bf v}_D \right ) \right ]	\ \ \ .
	\label{47}
\eeq
\noindent Identifying with the parton model result (\ref{46}), and chosing
${\bf v}_D$, ${\bf v}_B$, ${\bf v}^2_D$ and $w$ as independent variables
(only $w$ is covariant), we find the identity~:
\beq
\left ( {\bf v}_D \cdot {\bf v}_B \right ) \left [ X(w) + 2T_{1/2}(w) - 2
\right ] +
{\bf v}^2_D \left [ X(w) - 2T_{1/2}(w) \right ] + (w + 1) \left [ X(w) - 1
\right ] = 0
\label{48}
\eeq
\noindent that implies the solution (\ref{38}). \par

By the way, notice that formula (\ref{44}) leads to a simple computation of the
r.h.s. of
(\ref{10}). Particularizing (\ref{44}) to $(V^{\mu} - A^{\mu}) (V^{\nu} -
A^{\nu})$,
i.e. $\Gamma^{sc} = \gamma^{\mu} (1 - \gamma_5)$, $\Gamma^{cb} = \gamma^{\nu}
(1 -
\gamma_5)$, we get
\newpage
\[ <0|J^{sc}(0) \widetilde{J}^{cb}({\bf q})|B({\rm v}_B)> =
{f_B \sqrt{m_B} \over {\rm v}_D^0 \sqrt{2 {\rm v}_B^0}} \cdot \]
\beq
\left [ {\rm v}_D^{\mu} {\rm v}_B^{\nu} + {\rm v}_D^{\nu} {\rm v}_B^{\mu} -
g_{\mu \nu}
{\rm v}_D \cdot {\rm v}_B - i \varepsilon^{\mu \nu \rho \sigma} {\rm v}_{D\rho}
{\rm v}_{B\sigma} \right ] \label{49bis}  \eeq
\noindent  Contracting with the same tensor except for ${\rm v}_c \to {\rm
v}'_c = {m_b
{\rm v}_b - m_c {\rm v}_c \over m_c}$ and using ${\rm v}_b \cdot {\rm v}_c =
{\rm v}_b
\cdot {\rm v}'_c = w$, ${\rm v}_c \cdot {\rm v}'_c = 2w^2 - 1$ we obtain
${f_B^2 m_B
\over 2({\rm v}_D^0)^2 {\rm v}_B^0} 4w^2 = 2f_B^2 m_B$ in the $B$ rest frame.
{}From
(\ref{18}) we then obtain the result (\ref{10}). \par

In particular, the equation $X(w) = 1$ implies a formula for the slope of the
elastic IW
function. Deriving this equation at $w = 1$, we get~: \beq \rho^2 = \sum_{n
\not= 0}
{f^{(n)} \over f^{(0)}} \ \xi^{(n)'}(1) \ \ \ . \label{49} \eeq \noindent We
get a new
formula for the slope in terms of radial excitations alone. This is to be
compared to BIW
formula (\ref{4}) that expressed the slope in terms of ground state to $P$-wave
matrix
elements. We must add that (\ref{49}) does not have the positivity properties
of (4) and
does not imply any lower bound for $\rho^2$. Notice that the sum rules
(\ref{38}),
(\ref{49}) imply drastic cancellations among the ground state and the excited
states.
To have a feeling of what happens, we can compute $f^{(n)}$ and $\xi^{(n)}(w)$
in the
non-relativistic harmonic oscillator model. One gets for $\omega \cong 1$
\bea
&&\xi(w) \cong 1 + {R^2m^2 \over 2 \sqrt{2}} (w - 1) \nn \\
&&{f^{(n)} \over f^{(0)}} = (-1)^n \left [ {(2n+1)!! \over 2^n n!} \right
]^{1/2} \nn \\
&&\xi^{(n)} (w) \cong {(-1)^n \over n !} \left [ {n ! \over 2^n (2n + 1)!!}
\right
]^{1/2} \left ( {R^2m^2 \over \sqrt{2}} \right )^n (w - 1)^n \ \xi (w)
\eea
\noindent where $R$ is the radius of a light $q\bar{q}$ meson and $m$ is the
constituent
$q$ mass. One sees that (\ref{49}) is satisfied by the contribution of the
first radial
excitation, since higher excitations vanish as $(w - 1)^n$ $(n \geq 2)$. \par

The sum rules that we have found in the heavy quark limit have a simple,
explicit and
easy proof in the non-relativistic limit (i.e. if the light quark were also
non-relativistic). Indeed, we find, for a non-relativistic two-body system~:
\beq
\sum_{n} \psi_n ({\bf 0}) < n|{\bf r}^2|0 > = 0					\label{50}
\eeq
\noindent which is the non-relativistic version of (\ref{49}). The
demonstration follows
from  \beq
\sum_{n} \psi_n({\bf 0}) \int d {\bf r} \ \psi_n^*({\bf
r}) \ {\bf r}^2 \ \psi_0({\bf r}) = \int d{\bf r} \ {\bf r}^2 \ \delta ({\bf
r})
\ \psi_0({\bf r}) = 0	 \label{51}
\eeq
\noindent since
\beq
\sum_{n} \psi_n({\bf r}) \ \psi_n^*({\bf r'}) = \delta ( {\bf r} - {\bf r'})
\label{52}
\eeq
\noindent Similarly, one can generalize for any value of the recoil ${\bf q}$~:
\beq
\sum_{n} \psi_n({\bf 0}) < n|e^{i{\bf q} \cdot {\bf r}}|0 > = \psi_0({\bf 0})
\label{53}
\eeq
\noindent and one can deduce, similarly~:
\beq
\sum_{n} {\nabla} \psi_n({\bf 0}) \cdot < n|{\bf r} \ e^{i{\bf q} \cdot {\bf
r}}|0 >
= 3 \psi_0({\bf 0})			 \label{54}
\eeq
\noindent Equations (\ref{53}) and (\ref{54}) are the non-relativistic limit
version of
our sum rules in the heavy quark limit (\ref{38}). In (\ref{53}) and (\ref{54})
only
respectively the $S$ and $P$ waves contribute to the sum over the spectrum. In
(\ref{53}), the wave functions at the origin $\psi_n(0)$ are related to
$f^{(n)}$ and one
recovers $X(w) = 1$ (\ref{38}) for $w \cong 1$ (non-relativistic
approximation). In the
Pauli approximation (first relativistic correction) the gradient at the origin
${\nabla}
\psi_n({\bf 0})$ is, introducing spin, related to $f_{1/2}^{(n)}$ and
$f_{3/2}^{(n)}$.
In the limit in which one of the quarks has infinite mass, $f_{3/2}^{(n)} = 0$
and one
recovers the first equation (\ref{38}), $2T_{1/2}(w) = 1$, for $w \cong 1$.
\par

In conclusion, from duality arguments applied to the non-leptonic quantity
$\Delta
\Gamma$, or to the matrix element of the product of two currents between the
vacuum and
a pseudoscalar heavy meson, we have found new sum rules that involve the
Isgur-Wise
functions $\xi^{(n)}(w)$ and $\tau^{(n)}_{1/2}(w)$ and the decay constants
$f^{(n)}$,
$f^{(n)}_{1/2}$  ($n$ denotes any radial excitation). These sum rules give a
new
insight on the role of the excitations of the spectrum in the heavy quark
theory, and on
their relation to the slope of the elastic Isgur-Wise function. \par
\vskip 5 truemm

\noi {\bf Acknowledgements} \par
This work was supported in part by the CEC Science Project SC1-CT91-0729 and by
the
Human Capital and Mobility Program, contract CHRX-CT93-0132.

\vskip 5 truemm
\noindent{\fourteenbf Figure Captions} \bigskip
\begin{description}
\item{\bf Fig. 1.} The cut of the box diagram corresponding to the exchange
diagram.
\vskip 3 truemm

\item{\bf Fig. 2.} The cut of the box diagram corresponding to the spectator
diagram.

\vskip 3 truemm

\item{\bf Fig. 3.} Generic final states $D_s\bar{D}_s$, common to $B_s$ and
$\bar{B}_s$,
from the color allowed spectator diagram.
 \vskip 3 truemm

\item{\bf Fig. 4.} The matrix element between $\bar{B}_s$ and vacuum of the
product of
currents $A^{sc}_0(0)\widetilde{V}^{cb}_0({\bf q})$, estimated within the
parton model or by a sum over bound states in the heavy quark limit.   \vskip 3
truemm
\end{description}


\begin{thebibliography}{11}
\bibitem{[note]} We will use throughout the paper the quark model notation~:
$S$-wave mesons mean $J^P = 0^{-}, 1^{-}$ states and $P$-wave, $J^P = 0^+, 1^+,
2^+$.
\end{thebibliography}
 \end{document}